\documentclass[fleqn,10pt]{wlscirep}

\usepackage{enumerate}
\usepackage{graphicx}
\usepackage{epsfig}
\usepackage{color}
\usepackage{hyperref}
\usepackage{bm}

\title{Chemical Oscillation in Ultracold Chemistry}

\author[1]{Subhrajit Modak\thanks{corresponding author}}
\author[2]{Priyam Das}
\author[3]{Challenger Mishra}
\author[1]{Prasanta K. Panigrahi}
\affil[1]{Indian Institute of Science Education and Research Kolkata, Mohanpur, West bengal - 741246, India}
\affil[2]{Department of Physics, Amity Institute of Applied Physics, Amity University, Kolkata - 700156, India}
\affil[3]{Christ Church, University of Oxford, Oxford OX1 1DP, United Kingdom}
\affil[1]{modoksuvrojit@gmail.com}

\begin{abstract}
We demonstrate the occurrence of oscillatory reactions in the ultra-cold chemistry of atom-molecular Bose-Einstein condensate. Nonlinear oscillations in the mean-field dynamics occur for a specific range of elliptic modulus, giving rise to both in- and out-phase modulations in the atom-molecule population density. The reaction front velocity is found to be controlled by photoassociation, which also regulates the condensate density. Two distinct pair of in-phase bright localized gap solitons are found as exact solutions, existence of one of which necessarily requires a background. Cnoidal atomic density-waves in a plane wave molecular background are observed in both attractive and repulsive domains. Role of intra- and inter-species interactions on both existence and stability is explicated in the presence of photoassociation.
\end{abstract}

\begin{document}

\flushbottom

\maketitle

\section*{Introduction}
After the experimental realization of atomic Bose–Einstein condensate (BEC)\cite{eo,eo1} at nanokelvin temperatures, a major research effort over the past few years has been to extend the techniques of atom cooling and trapping\cite{ac,bc} to molecular systems for realizing molecular Bose-Einstein condensates. Complex spectral structure of molecules has made it difficult to cool them to the ultracold regime by direct laser-cooling techniques that has successfully worked for atoms.  Although significant progress has been made for capturing molecules in different opto-magnetic trapping configurations\cite{mot1,mot2}, such techniques have not been successful in preparing dense samples of molecules in specific quantum states. An alternative pathway has now been followed for realizing molecular condensates through the conversion of pre-cooled atomic BECs\cite{1,2,3,4}. This approach successfully exploits the existence of scattering resonances\cite{fr,fr1} for connecting ultracold atoms to transient resonant states. For example, a two-photon
stimulated  Raman  transition  in  a ${}^{87}\text{Rb}$ BEC  has  been  used  to
produce ${}^{87}\text{Rb}_{2}$  molecules in a single rotational-vibrational state\cite{tw3}. Ultracold molecules have also been formed through photoassociation (PA)\cite{tw4,tw5}. An immediate advantage of atom-molecule co-trapping is that it offers longer order of interaction time compared to the molecular crossed beam methods.  This can facilitate the study of cold chemistry for especially slow processes. The  prospect  of creating superposition  of  atomic  and  molecular  condensates has  initiated  much  theoretical  work\cite{tw1,tw2}, although the coherence properties of  these  systems  have  not been probed in great detail. Experimentally PA can be effectively used to produce  coherent  coupling  between  atomic  and  molecular BECs for investigating the atom-molecular interaction within the life time of trapped molecules.
\\

\begin{figure}[]
\begin{center}
\includegraphics[scale=0.5]{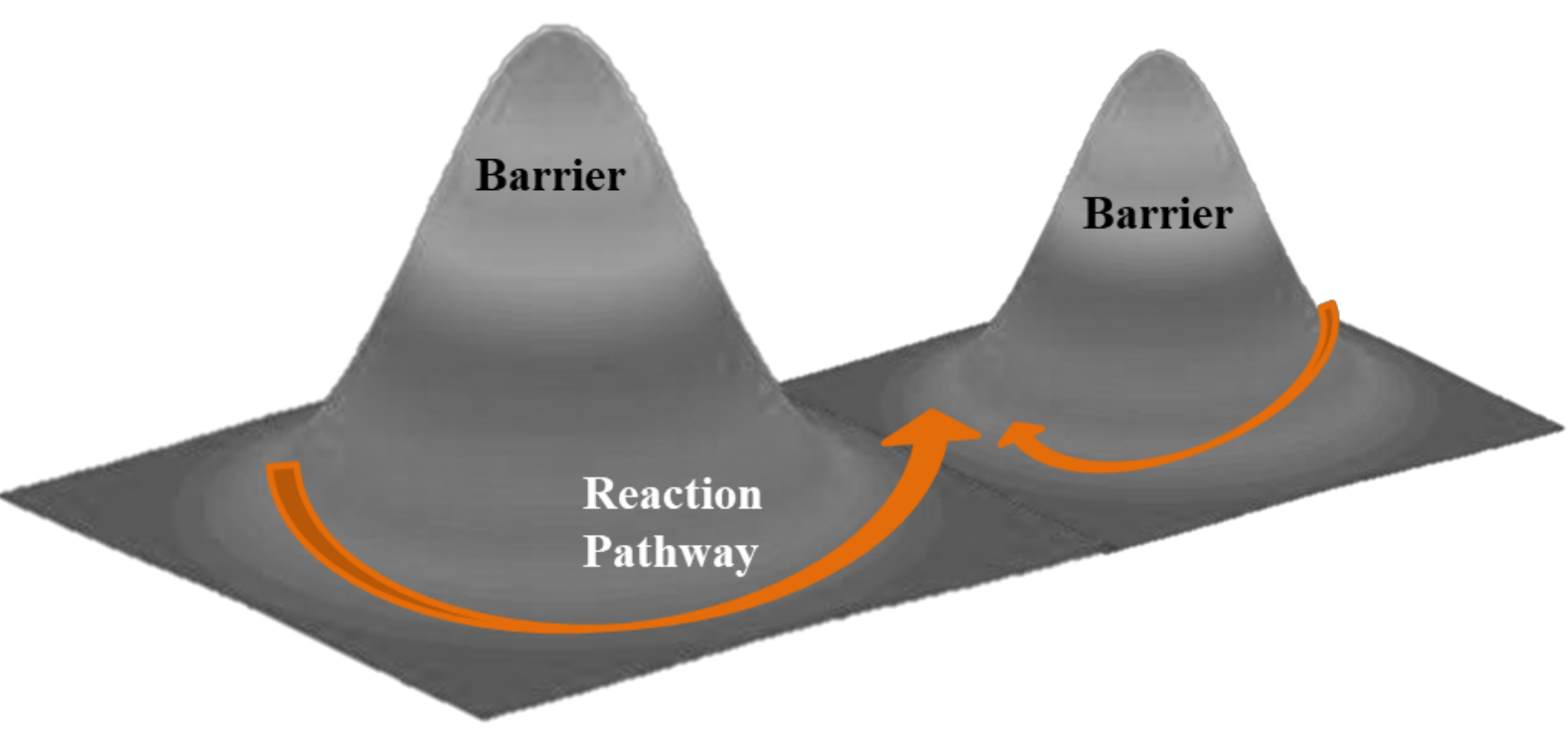}
\caption{Schematic representation of reaction pathways in ultracold reaction, avoiding the effective barrier in conventional reactions.} 
\label{po}
\end{center}
\end{figure}

Chemical reaction at ultracold temperature can be surprisingly efficient, aided by nonlinear scattering effects. At such temperature, disordered movement of high momenta particles is absent.  Hence, conventional description\cite{7} of collision dependent reaction, based on Maxwell-Boltzman statistics gets replaced with the framework of quantum statistics, where reactants are characterized by their de-Broglie wavelengths. The first phenomenological model in this direction was proposed by Heinzen et al.\cite{5}, where a mean-field ansatz was used to describe the coherent formation of diatomic molecules in BEC. In this approach, reactants are represented by their corresponding fields, while the density operator replaces reactant concentration. The mean field description for coupled atom-molecule
BEC\cite{oles,tt} (AMBEC) has been developed, which includes pair
correlations,  quantum   fluctuations  and  thermal  effects. Dynamics of AMBEC is described  by  a  modified  coupled  non-linear  Schr$\ddot{o}$dinger equation (NLSE). The difference from pure two-species BEC\cite{aa,bb}, described by a coupled NLSE, arises due to inter-conversion, which induces quadratic non-linearity in addition to the cubic  non-linearity arising from s-wave scattering. Simultaneous appearance of cubic and quadratic non-linearity in AMBEC provides novel cross-phase modulation\cite{ca}, affecting the conversion process in the presence of PA. Fig\ref {po} schematically depicts the reaction pathways between atoms and molecules at ultraccold temperature, that can circumvent the conventional chemical barriers. In the mean field approach, suitable ground state solutions can asymptotically connect two different configurations, without reaching the top of the potential barrier. The complex nature of the mean field wave function enables this phenomenon, which is a manifestation of coherence, a prime example being the Lieb mode in BEC\cite{lie}.
\\
\\
 The present work is devoted to the study of oscillatory chemical reactions\cite{bm1,bm}  in the atom-molecular system. The role of various non-linear interactions and that of PA will be probed in detail. Oscillatory kinetics refers to spontaneous progression of reaction in both forward and backward directions, appearing to violate the second law of thermodynamics. These reactions occur far from thermodynamic equilibrium\cite{co} and do not last forever, dying away slowly as the mixture settles into an unchanging state. In the present case, nonlinear oscillations are found to set in for restricted values of elliptic modulus, giving rise to both in- and out-phase modulations in the atom-molecule population density. PA is found to play the key role in the oscillatory reaction, controlling the speed of the reaction front as well as its amplitude. We find the exact parametric conditions, separating different non-linear excitations. Exact localized solutions are found to be in-phase and gapped\cite{tup}, differing significantly from the two-BEC\cite{pup} case, with one class of soliton necessarily accompanied by a background. Interestingly, nonlinear excitations in the form of \emph{cnoidal} waves, similar to the ones in atomic BEC, for both the repulsive and attractive domains, are found as exact solutions, wherein the heavy molecular component has a plane wave character. This is similar to the case of optical fiber, with a core and clad component, core allowing solitonic excitations, whereas the clad supports plane wave modes\cite{tsr}. 

\section*{Model}
Two atoms can be combined together to form a molecule through the absorption of a photon from an applied optical field during an atomic collision.  In recent years, Feshbach resonances\cite{fbr1,fbr2} have come into prominence in the study of ultracold atomic gases, wherein positions of resonances can be adjusted using applied magnetic fields. It is possible to control
the interactions between atoms and molecules appropriately by tuning resonances to near-zero collision energy. This interaction close to the absolute zero of temperature has been christened as \emph{ultracold chemistry}\cite{uc1,uc2,uc3,uc4}. Since the energy produced in this exoergic process is very low, the reaction products remain in the trap. We are considering only the elementary reactions that proceed without forming any identifiable intermediate species. The most elementary second-order reaction, \emph{diatomic molecule formation}, is represented by:
\begin{center}
$A + A \overset{k}{\rightleftarrows} A_2$.
	\label{eq:molecule}
\end{center}
In this case, possible product formation leads to different possibilities  for  the  quantum statistics: $bb\rightarrow b$, $bf\rightarrow f$ and $ff\rightarrow b$, where $b$ stands for bosonic and $f$ for fermionic counterpart.
Interestingly, these conversions correspond to well-known field theroretical models, the  Lee-Van  Hove model  of  meson  theory,  and  the  Friedberg-Lee  model  of high-$T_{C}$ superconductivity\cite{po}. We consider a  chemical  system  of  the  first  type, where  bosonic  enhancement of the chemical dynamics is the strongest.  
\begin{table}[h!]
	\centering
		\begin{tabular}{c|c|c}
		 Order&Reaction & $\hat{H}_{\text{int}}$ \\ \hline
			0.&$\text{Bath} \overset{k}{\rightleftarrows} A$ & $k(\hat{a}_A^\dag+ \text{h.c.})$  \\
			1.&$A \overset{k}{\rightleftarrows} B$ & $k(\hat{a}^\dag_{A} \hat{a}_{B} + \text{h.c.})$  \\
			2.&$A + A \overset{k}{\rightleftarrows} A_{2}$ & $k(\hat{a}_{A}^\dag\hat{a}_{A}^\dag\hat{a}_{A_2} + \text{h.c.})$  \\
			
		\end{tabular}
		\vspace{3mm}
		\caption{The proposed interaction Hamiltonians\cite{7} for low-order bosonic reactions.}
			\label{tab:interaction}
\end{table}
Table \ref{tab:interaction} illustrates different orders of chemical reaction in a general scenario.  Zeroth-order reaction  physically represents an exchange of species with reservoir, while the First-order reaction models a linear interaction between two quantum fields. On the other hand, atom-molecule inter-conversion requires a Hamiltonian involving second-order reaction: 
\begin{equation}
\hat{H} =E_{A}\hat{n}_A +E_{A_2}\hat{n}_{A_2} + k(\hat{a}_A^\dag\hat{a}_A^\dag\hat{a}_{A_2} + h.c.),
		\label{eq:quantum_hamiltonian}
\end{equation}
which can be generalized to include the effect of multiple concurrent reactions, particle loss and dissipation. For the rest of the paper, we neglect these effects and assume the reaction rate to be much larger than the ground state energies, i.e., $k \gg |E_A| + |E_{A_2}|$, where $E_A$ and $E_{A_2}$ label the corresponding ground-state energies. Notice that, within this framework, we are restricted to the description of reversible reactions with a single reaction rate. Moreover, this model can only probe the outcome of a chemical reaction, providing no direct information regarding the actual process of bond breaking and bond making. The Hamiltonian for such system can be written in terms of field operators for atoms and for the molecular resonant state. At resonance, the number of molecules becomes considerable and molecular BEC is formed. We  take  into  account  two  body  atom-atom, molecule-molecule and atom-molecule collisions, as well as the term responsible for transfer of pairs of atoms into molecules and vice versa in the unit $m_a=1$:
\begin{eqnarray}
	\hat{H} &=& \large{\int} {\rm d}^3 r \Big( \hat{\psi}^{\dagger}_a \Big[ -\frac{\hbar^2}{2}\nabla^2 
	+ V_a(\vec r) + \frac{g_a}{2} \hat{\psi}^{\dagger}_a \hat{\psi}_a \Big] \hat{\psi}_a \nonumber \\
	&+& \hat{\psi}^{\dagger}_m \Big[ -\frac{\hbar^2}{4}\nabla^2  
	+ V_m(\vec r) + \epsilon + \frac{g_m}{2} \hat{\psi}^{\dagger}_m \hat{\psi}_m \Big] \hat{\psi}_m \cr \nonumber \\
	&+& g_{am} \hat{\psi}^{\dagger}_a  \hat{\psi}_a  \hat{\psi}^{\dagger}_m  \hat{\psi}_m
	+ \frac{\alpha}{\sqrt{2}} \big[ \hat{\psi}^{\dagger}_m \hat{\psi}_a \hat{\psi}_a 
	+ \hat{\psi}_m \hat{\psi}^{\dagger}_a \hat{\psi}^{\dagger}_a \big] \Big)
	\label{HFR}	
\end{eqnarray}
Here $g_{a}$, $g_{m}$ and $g_{am}$ measure the strength of atom-atom, molecule-molecule and atom-molecule interactions respectively. $V_a$ and $V_m$ stand  for  atomic  and  molecular  trapping  potentials, with $\alpha$ being the strength of PA. The parameter $\epsilon$ is the energy mismatch in converting the atoms to molecules. In the following we consider a cigar shaped geometry. The modified parameters in the case of quasi one dimensional geometry\cite{sala,sala1} are kept unchanged for notational convenience. Equations of motion for the atomic and molecular mean fields, are given by
\begin{eqnarray}
i\frac{\partial\psi_{a}}{\partial t} &=& -\frac{1}{2}\frac{\partial^{2} \psi_{a}}{\partial x^{2}} + (V_a+g_{a}|\psi_{a}|^{2} + g_{am} |\psi_{m}|^{2}) \psi_{a}  + \alpha \sqrt{2}\psi_{m}\psi_{a}^{*}, \label{ac} \nonumber\\ \\
i\frac{\partial\psi_{m}}{\partial t} &=& -\frac{1}{4}\frac{\partial^{2} \psi_{m}}{\partial x^{2}}+(V_m+\epsilon +g_{m}|\psi_{m}|^{2}+g_{am}|\psi_{a}|^{2})\psi_{m} + \frac{\alpha}{\sqrt{2}}\psi_{a}^{2} \label{mc}. \nonumber\\
\end{eqnarray}
Evidently,  for nonzero $\alpha$, the overall particle number is conserved,
\begin{equation}
N=\int{(\vert\psi_{a}\vert^2+2\vert\psi_{m}\vert^2}) dx=N_{a}+2N_{m}.
\end{equation}
Here $\psi_{j}$'s are taken as, $\psi_{j}(x,t)=\sqrt{n_{j}(x,t)}e^{i\phi_{j}(x,t)}$ for $j=a,m$., for which the continuity equation can be written as,
\begin{equation}
\frac{\partial}{\partial {t}}(n_{a}+2n_{m})+\frac{\partial}{\partial {x}}(\sum_{a,m}n_{j}\frac{\partial\phi_{j}}{\partial x})=0.
\end{equation}
This condition is invariant under scale transformation and Galilean boost\cite{boo}. Under scaling, density and phase change as,   $n(x,t)\rightarrow{\beta n(\beta x,\beta^2 t)}$, $\phi(x,t)\rightarrow{\phi(\beta x, \beta^2 t)}$, while for the boost by an amount $v$, the changes are $n(x,t)\rightarrow n(x-vt,t)$ and $\phi(x, t)\rightarrow\phi(x-vt, t)+v[x-vt/2]$. In the following, we first consider the static configurations, which yield the asymptotic equilibrium states with given initial configurations of atoms and molecules and highlight some compatible excitation pairs that induce desired reactions. One needs to consider the trapping potential for possible comparison with experimental scenario.

\section*{Trapping Configuration}
Confining  traps  are usually approximated  by harmonic potentials. Trap frequency in general, can be time-dependent. Depending on the sign of trap frequency $\omega(t)$, oscillator potential can either be confining or expulsive. Interestingly, in the mean field approach, the wave packet dynamics in presence of a time-dependent trap can be related to the dynamics without a trap through the similarity transformation\cite{gsa}:
\begin{equation}
\psi_j(f,g)=M_j(t)\psi_j\Big[f(x,t)\Big]e^{i\phi(x,t)}
\end{equation}
Here $M(t)$ represents the amplitude of the pulse and $f(x,t)$ is the similarity variable
\begin{equation}
f(x,t)=\frac{x-x_c}{w(t)}    
\end{equation}
where $w(t)$ and $x_c$ are the dimensionless width and center of the self-similar wave.  The quadratically
chirped phase is given by
\begin{equation}
\phi(x,t)=c(t)\frac{x^2}{2}+b(t)x+a(t)    
\end{equation}
where the parameters $c(t)$, $b(t)$ and $a(t)$ are to be determined. They are related to the phase-front curvature, the frequency shift, and the phase offset, respectively. In the general case of time-dependent harmonic trapping, condensate profile gets appropriately modulated in time, along with the conditions: 
\begin{eqnarray}
\dot{c}-c^2(t)=\omega(t), \\
\ddot{x_c}+\omega(t)x_c=0.
\end{eqnarray}
The first one, involves the  motion related to chirping, which can be expressed as a Schr$\ddot{o}$dinger eigenvalue problem via the Cole-Hopf transformation\cite{cole}. Taking advantage  of  this  connection, it can be shown that, corresponding to each solvable quantum-mechanical system, one  can  identify  a  soliton  configuration. The  fact  that  the Schr$\ddot{o}$dinger equation can be exactly solved for a variety of potential, gives us freedom to control the dynamics of the BEC in a  number  of  analytically  tractable trap configuration.  This type of chirped phase regularly arises in nonlinear fiber optics, as an acceleration induced inhomogeneity, which can balance the effect of harmonic trap. On the other hand, the later condition is a manifestation of $\emph{Kohn}$ mode, depending solely on the frequency of the harmonic trap. The time-dependent trap and scattering length can be used to compress and accelerate the wave, leading to the possibility of their coherent control\cite{nath}.

 \section*{Ground state configuration} \label{sgs}
\begin{figure}[t]
\begin{center}
\includegraphics[scale=0.28]{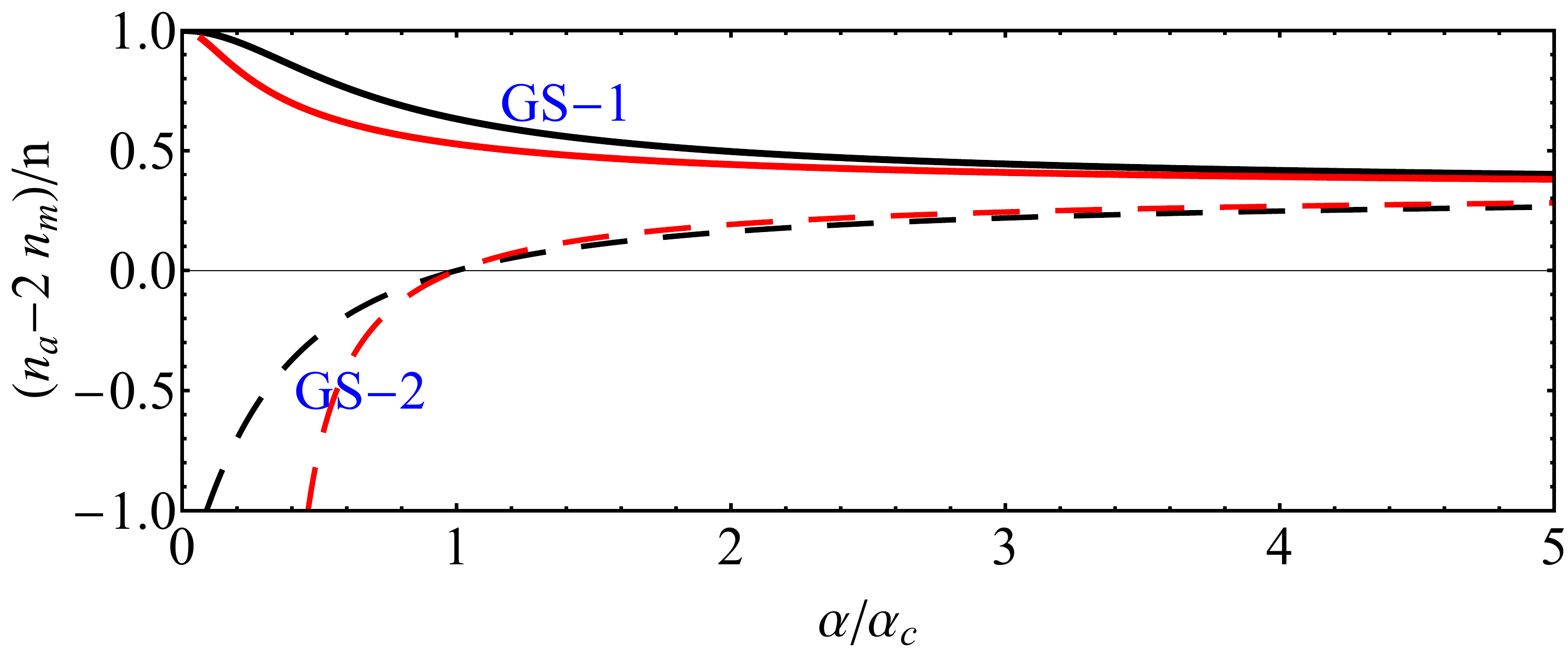}
\caption{Two different ground states (GS-1 and GS-2) are shown for the initial conditions corresponding to the presence and absence of photoassociation for two different energy mismatch values.}
\label{fig2}
\end{center}
\end{figure}

The ground state is governed by the values of densities and phases that minimize the energy per unit volume. Energy density for the case of constant density and phase, assuming same chemical potential for both the condensate components, is given by

\begin{eqnarray}
  \mathcal{E} = \frac{1}{2} g_{a} n^{2}_{a} + \frac{1}{2} g_{m} n^{2}_{m} &+&  g_{am} n_{a} n_{m}+\epsilon n_{m} - \mu (n_{a} + 2n_{m}) \nonumber \\ &+& \frac{\alpha}{\sqrt{2}} n_{a} \sqrt{n_{m}} \cos(\phi_{am}).
\end{eqnarray}
Here, $\phi_{am}$ is the phase difference between the condensate components, $\phi_{am}=\phi_{m}-2\phi_{a}$, leading to a phase correlation in the presence of PA, different from the density-density correlations arising from inter-species interaction. As mentioned earlier, this term can arise from the two-photon (Raman) process or a direct Rabi coupling between the components. The minimum energy configuration corresponds to $\phi_{am} = \pi$, with the equilibrium configuration characterized by,
\begin{equation}
  (g_{a}-\frac{g_{m}}{4}) n+p ((g_{a}+\frac{g_{m}}{4})-g_{am})-\frac{\alpha}{2\sqrt{2}}\frac{n-3p}{\sqrt{n-p}}-\epsilon=0,
\end{equation}
where $p=n_{a}-2n_{m}$ is the density difference. For convenience, this state equation can be written as a cubic polynomial in $p/n$,
\begin{equation}
\Big(\frac{p}{n}\Big)^3-\big[1-2\Big(\frac{F}{G}\Big)-\frac{9\alpha^2}{8nG^2}\Big]\Big(\frac{p}{n}\Big)^2-\Big[2\Big(\frac{F}{G}\Big)-\frac{6\alpha^2}{8nG^2}-\Big(\frac{F}{g}\Big)^2\Big]\Big(\frac{p}{n}\Big)-\Big[\Big(\frac{F}{G}\Big)^2-\frac{\alpha^2}{8nG^2}\Big]=0, 
\end{equation}
 where, $F=(g_a-\frac{g_m}{4})$ and $G=((g_a+\frac{g_m}{4})-g_{am})$. For this configuration, we keep the mismatch part to zero. Generally, cubic equations of state are much used in thermodynamical system, arises due to the addition of a  co-volume parameter $b$ and an attractive pressure term inversely proportional to $V^2$ (molar volume) to the ideal gas equation. This form of equations are capable of representing different phases at temperatures below and above the critical point. In such cases, one will either have one or three real roots, depending on, whether the phases are co-existing or separated from others. Since we are dealing with physical quantities, only real roots are of interest. More specifically, we look for real, positive roots of $p/n$. We fix our case in a way such that $p/n$ could vary only within -1 to +1. At these points, the density is either occupied by atoms ($p/n=1$) or by molecules ($p/n=-1$). For any values of $p/n$ in between -1 to +1, the density is shared by the mixture of atoms and molecules. The structure of the ground state is better understood if the interactions are repulsive. For an example, we take $g_a=g_m=1/2$ and $g_{am}=1/4$. Fig (\ref{fig2}) represents numerically computed roots of Eq.[14] for different inter-conversion strength. The solid-black curve (GS-1) corresponds to the case, when $\alpha=0$, implying the condensate is having only an atomic part. On the other hand, the case (GS-2), shown by dashed-black curve, corresponds to finite inter-conversion to start with. As $\alpha$ starts to increase, the ratio between the density difference to the total density decreases. In other words, atomic density rises in comparison to the molecular proportion. In this process, when $\alpha$ comes to a certain critical strength, $\alpha_{c}=2\sqrt{2}[(g_{a}-g_{m}/4)\sqrt{n}]$,  the ratio between density difference to the total density becomes half. Further increment in $\alpha$ takes the system into a state, where both atoms and molecules co-exist, accompanied by a constant density difference. As will be evident later, the condensate configuration will no longer be stable upto this strength of inter-conversion. Here, critical strength does not mark any appearance of phase transition. It is about quantifying a particular ratio of $p/n$. The solid-red and dashed-red curves are explicated in the presence of a finite mismatch (positive) term, where the critical strength is reduced by a  factor of $2\sqrt{2}\epsilon/\sqrt{n}$.
\section*{Stability}
\begin{figure}[t]
\begin{center}
\includegraphics[scale=0.30]{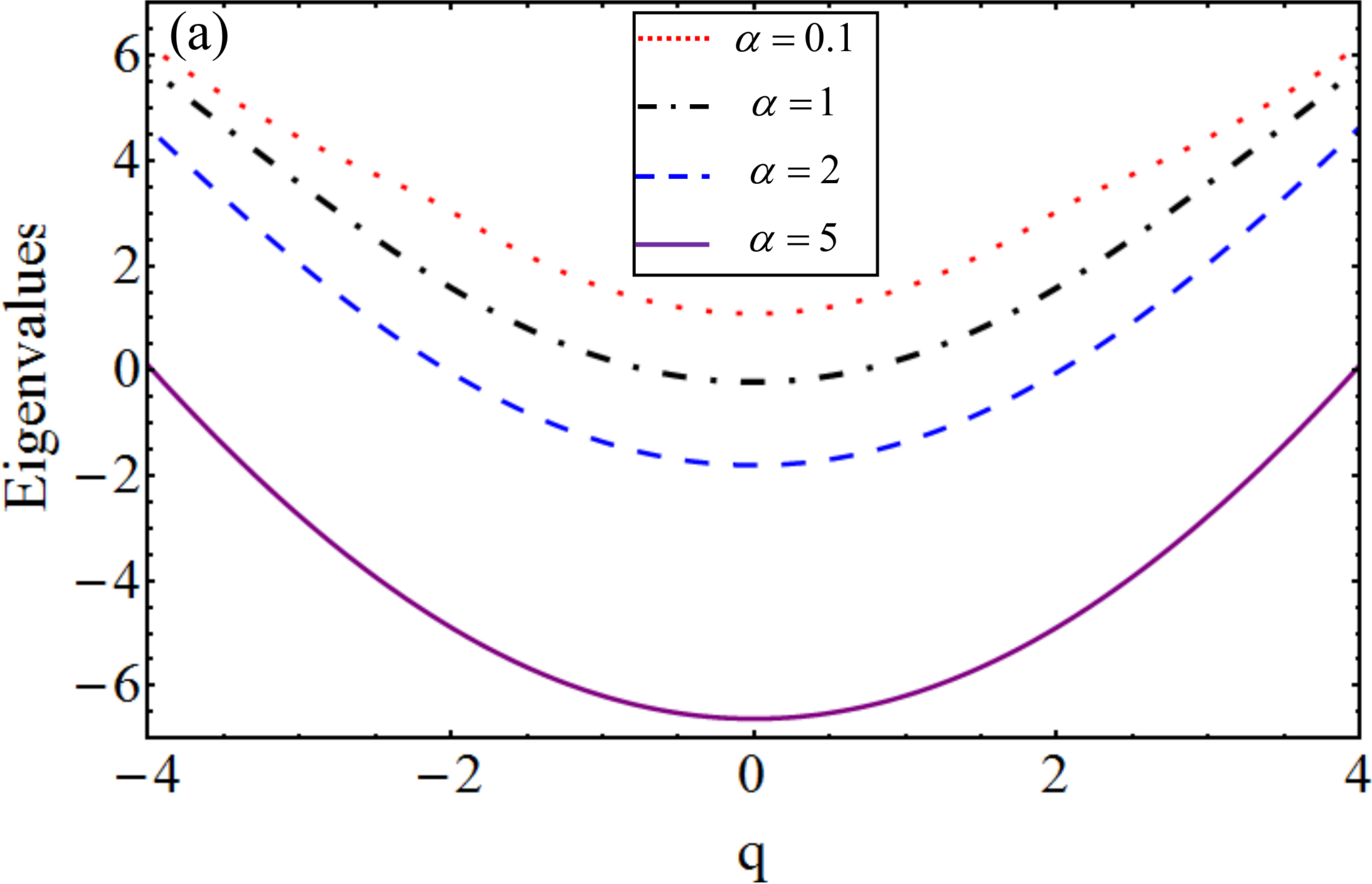}
\includegraphics[scale=0.31]{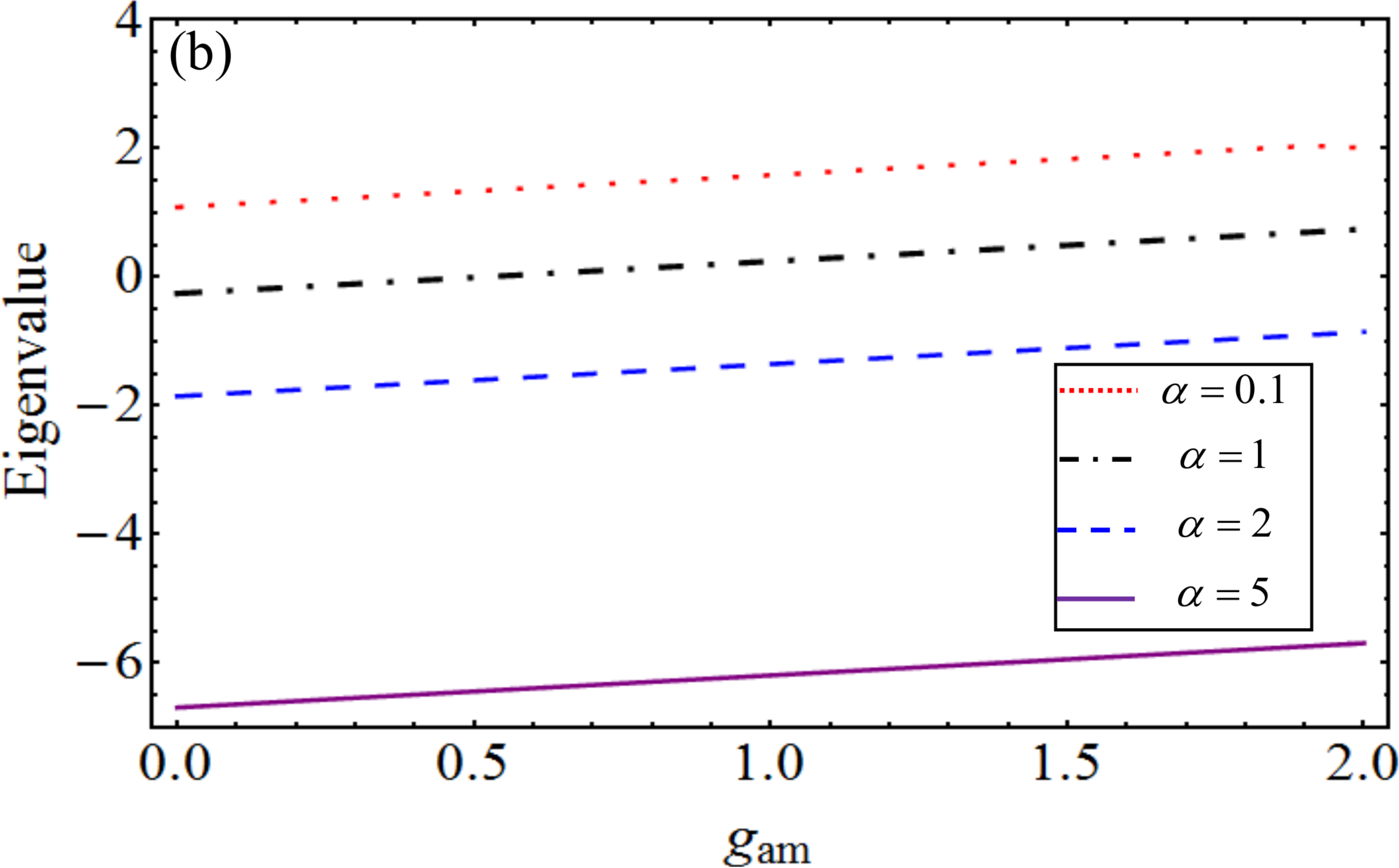}
\caption{ Depiction of the spatial modulational instability. (3a) shows the domain of negative eigenvalues, representing the region of instability, (3b) shows the weak dependence of eigenvalues on atom-molecule scattering effect.}
\label{fig3}
\end{center}
\end{figure}

We now examine the effect of small variation of the condensate density due to the presence of phonons, which can potentially destabilize the condensation. In case of a single species condensate, stability depends on the sign of interaction. For repulsive interaction, condensate is found to be stable, whereas for attractive case, it is unstable after a critical value of the interaction. In case of a two-species system, it has been observed that even if intra- and inter-species interactions are repulsive, instability may occur in the system\cite{stbl1}. To examine stability, we insert a weak perturbation of the form $\delta\tilde\phi_{j}\sim\delta\phi_j e^{i(qx-\omega t)}$ of frequency $\omega$ and wave vector $q$ around the steady state solution\cite{stbl}, 
\begin{eqnarray*}
  \phi_{a}(x,t) &=& (\phi_{a0}+ \delta\tilde\phi_{a}(x,t))e^{-i\mu t}, \\ \quad
  \phi_{m}(x,t) &=& (\phi_{m0} + \delta\tilde\phi_{m}(x,t))e^{-2i\mu t},
\end{eqnarray*}
where $\phi_{j0}$ are the real densities to start with and $\delta\tilde\phi_j$ satisfies $\delta\tilde\phi_{j}\ll\phi_{jo}$. Linearizng in $\delta\phi_j$, one obtains the dispersion relation. The corresponding gain/loss spectrum can be put into a matrix form $\mathcal{M}$, with the diagonal elements $\mathcal{M}_{11} = \frac{q^2}{2} + 2 g_{a} \vert\phi_{a0}\vert^2 + g_{am}\vert\phi_{m0}\vert^2 - (\mu+\omega)$, $\mathcal{M}_{22} = \frac{q^2}{4} + 2 g_{m}\vert\phi_{m0}\vert^2 + g_{am}\vert\phi_{a0}\vert^2 + (\epsilon-2\mu-\omega)$ and the off-diagonal elements$\mathcal{M}_{12} = g_{am} \psi^{*}_{m0} \psi_{a0} + \sqrt{2} \alpha \psi^{*}_{a0}$, $\mathcal{M}_{13} = g_{a}\phi_{a0}^2 + \alpha\sqrt{2}\phi_{m0}$ and $\mathcal{M}_{34}=g_{am}\phi_{m0}\phi_{a0}^{*}+\alpha\sqrt{2}\phi_{a0}$, $\mathcal{M}_{24}=g_{m}\phi_{m0}^2$. It is evident that, the atom-molecular system is stable if the imaginary part of the eigenvalues of the matrix $\mathcal{M}$ are positive and becomes unstable if the lowest eigenvalue of $\mathcal{M}$ is negative. Fig.(\ref{fig3}) delineates the stablity/instability regions. This is physically justified since the fluctuation in the form of plane waves must not decrease the energy of the system for the stable condensate. Semi-positive eigenvalues of $\mathcal{M}$ lead to real normal mode frequencies (collective excitations), which in turn ensures the stability of the system. We investigate the behavior of the eigenvalues of the matrix $\mathcal{M}$ as a function of the quasi-momentum $q$, for different values of atom-molecular inter-conversion term. It can be seen from the Fig.(\ref{fig3}a) that, for small values of $\alpha$, the system stays in a stable domain. The red (dotted) and black (dash-dotted) curves correspond to $\alpha = 0.1$ and $1$, where the eigenvalues lie in the positive region, indicating that the system is stable. Further increasing the value of $\alpha$, e.g., the case with $\alpha = 2$ and $5$, respectively shown in blue (dashed) and black (solid) curves, makes the system unstable, as the eigenvalues lie in the negative regime. Fig.(\ref{fig3}b) depicts the eigenvalues as a function of inter-species interaction ($g_{am}$) for the same values of $\alpha$, showing minimal effect of $g_{am}$ on stability. As will be evident later, the condensate wave-packets and localized solutions are primarily controlled by PA.



\section*{Oscillatory Excitations}
We now investigate the dynamics of AMBEC, concentrating on the possibility of chemical oscillations. Oscillatory excitations manifest in several chemical reactions, most well-known being the Belousov-Zhabotinsky reaction\cite{bz,bz1}, where the products exhibit periodic changes either in space or in time and give rise to remarkable spatio-temporal pattern\cite{bz3}. In the present case, both linear and quadratic $\emph{cnoidal}$ waves are found as exact solutions. The linear case is analogous to the one in atomic BEC, whereas the quadratic one is novel to the atom-molecular system. Its presence crucially depends on photo-association. We start with the quadratic excitation and without loss of generality, consider the following pair of $\emph{cnoidal}$  waves, for both the condensate components:
\begin{eqnarray}\nonumber
  \phi_{a}(x, t) &=& \left(A + B~ \textrm{cn}^{2}(\xi, m) \right) e^{i k x}, \\
  \phi_{m}(x, t) &=& \left(C + D~ \textrm{cn}^{2}(\xi, m) \right) e^{2 i k x},
\end{eqnarray}
here, $\xi = \beta (x - u t)$ with $u$ being the velocity. The cnoidal wave excitations exist only in the presence of a fast varying plane wave component: $e^{i k x}$. A lengthy calculation leads to the amplitudes of periodic pair,
\begin{eqnarray}
  B &=&  \pm\frac{D}{2}; \qquad D = - \frac{3 \sqrt{2} \hbar^{2} \beta^{2} m}{\alpha} \\
  A &=& \frac{C}{2} - \frac{ \alpha}{\sqrt{2} g_{a}}; \qquad C = \frac{3 \alpha}{2 \sqrt{2} g_{a}} - \frac{m (m - 1)}{2 (2 m - 1)} \pm \frac{1}{2} \sqrt{\frac{ \alpha^{2}}{2 g_{a}^{2}} + \left(\frac{m (m - 1)}{2 m - 1}\right)^{2}}
\end{eqnarray}

along with $\beta^{2}=\frac{2m_{a}}{3 \hbar^{2}(1-2m)} \left( \frac{-3 \alpha}{2 \sqrt{2}}C +\frac{ 9 \alpha^{2}}{8 g_{a}}\right)$ and the wave vector, $k^{2} =-\frac{2\alpha^2}{g_{a}}$. 
It is seen that the nature of atom-atom interaction decides the sign of energy mismatch, $\epsilon=\frac{15\alpha^2}{8g_{a}}$, with $g_{a}=16g_{m}$ and $g^{2}_{am} = g_{a} g_{m}$. We consider $\epsilon<0$\cite{oles} and obtain explicit solutions for general values of couplings. Accessible density parameters are ensured for $g_{a},g_{m}<0$. PA plays crucial role in determining the front velocity\cite{cs} and can be used to control how quickly the reactants are used up. This is in sharp contrast to the prediction of usual chemical kinematics, where rates do not depend on the number of participating particles and tend to zero at low temperature. The parameter controlling PA, leads  to  two  different  physical  situations  in case of $\alpha=0$ and $\alpha\rightarrow 0$. In the absence of $\alpha$, velocity of excitation remains a free parameter, whereas for $\alpha\rightarrow 0$, the velocity tends to zero. Reaction rates at these temperatures can be made comparable to or even larger than they are at room temperature. It is evident that, $\emph{cnoidal}$ oscillation for both the components would exist only if $\frac{1}{2}< m \leq 1$. This pair of excitations lead to an unique elevation-depression or depression-depression density profile. On the other hand, no trigonometric counterpart can be found, as in the limit $m = 0$, amplitude vanishes, leaving only a constant background. This is different from the case of two BEC where sinusoidal excitation\cite{pup} appears as an exact solution. Densities of atomic and molecular BECs are explicated in Fig.(\ref{fig4}). Both the densities are characterized by two-frequency modulations.    
\subsection*{Cnoidal chemical waves in a background}
Unlike the quadratic oscillatory excitations for both the components, the AMBEC system also exhibits periodic atomic density waves in a constant molecular background,
\begin{eqnarray}\nonumber
  \phi_{a}(x,t) &=& \phi_{a0}~\textrm{cn}(\xi,m)e^{ikx},\\ 
  \phi_{m}(x,t) &=& \phi_{m0}~e^{2ikx}\nonumber \\ \label{an}
\end{eqnarray}
with $\xi=\beta(x-vt)$. Amplitudes of the atomic and molecular densities are found to be of the form, $\phi_{m0}=\frac{\alpha}{\sqrt{2} g_{am}}$, $\phi_{a0}^2=-\frac{m\beta^2}{2 g_{am}}$
with $\beta^4=\frac{8\alpha^2(\alpha^2+\epsilon g_{am})}{(9-17m)g_{am}}$, $k^{2} =v^2=-(\epsilon+\frac{g_{m}\alpha^2}{2g_{am}^2})$ and $g_{a}=2g_{am}$. Physical solutions are assured for $g_{a}, g_{m}, g_{am}<0$, implying both intra and inter-species interactions must be attractive. Interesting to note that,  this class of solution exhibits two disjoint domains: for $\frac{1}{2}<m<1$, energy mismatch $\epsilon < \frac{\alpha^2}{\vert g_{am}\vert}$ with $\vert g_{m}\vert>\frac{2}{\kappa}\vert g_{am}\vert$, while for the remaining half, $0<m\leq\frac{1}{2}$, with $\epsilon > \frac{\alpha^2}{\vert g_{am}\vert}$ and $\vert g_{m}\vert>2\kappa\vert g_{am}\vert$, where $\kappa$ is a positive number. If one considers atomic condensate in terms of elliptic $\textrm{sn}$ function, keeping the molecular density constant, one obtains, $\beta^2=\frac{(\frac{\alpha^2}{g_{am}}-v^2)}{(m+1)}$ and $\frac{\alpha^2}{g_{am}} > v^{2}$, with all other parameters remain same. It is  evident that, both intra- and inter-species interactions must be repulsive, $g_{a}, g_{m}, g_{am}>0$, this refers to a complete opposite interaction landscape.

\begin{figure}[t]
\begin{center}
\includegraphics[scale=0.39]{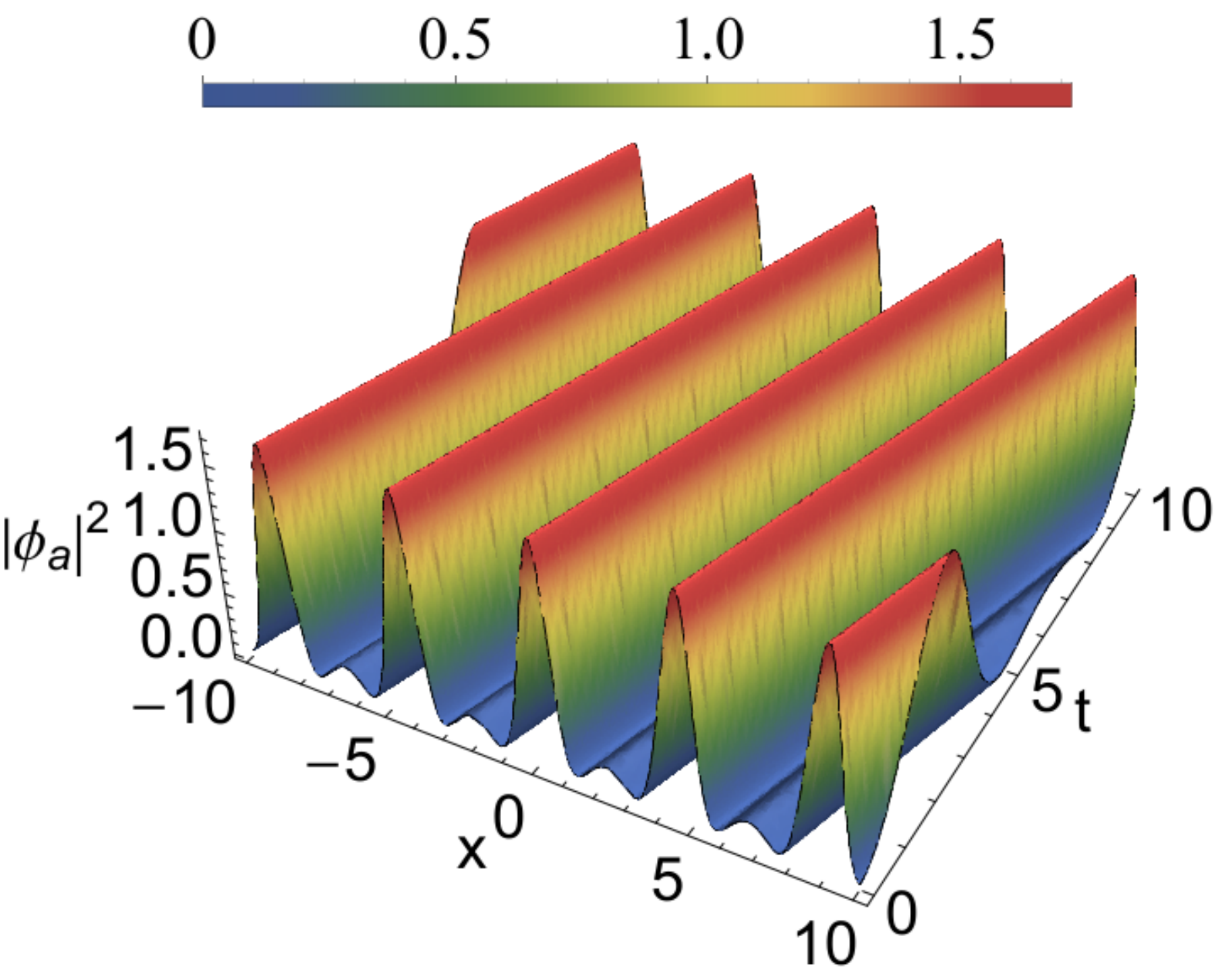}
\includegraphics[scale=0.39]{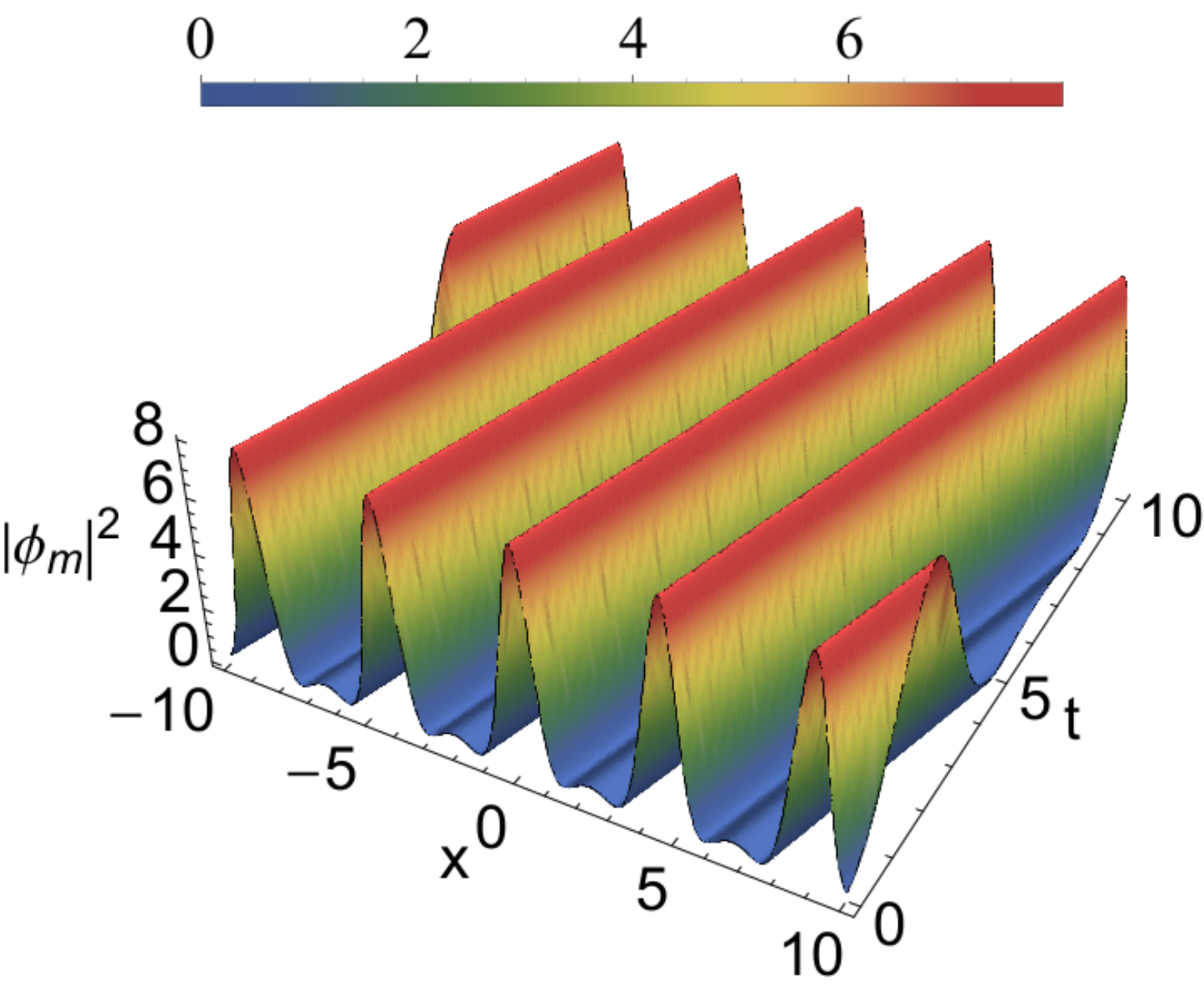}
\label{}
\end{center}
\end{figure}

\begin{figure}[t]
\begin{center}
\includegraphics[scale=0.28]{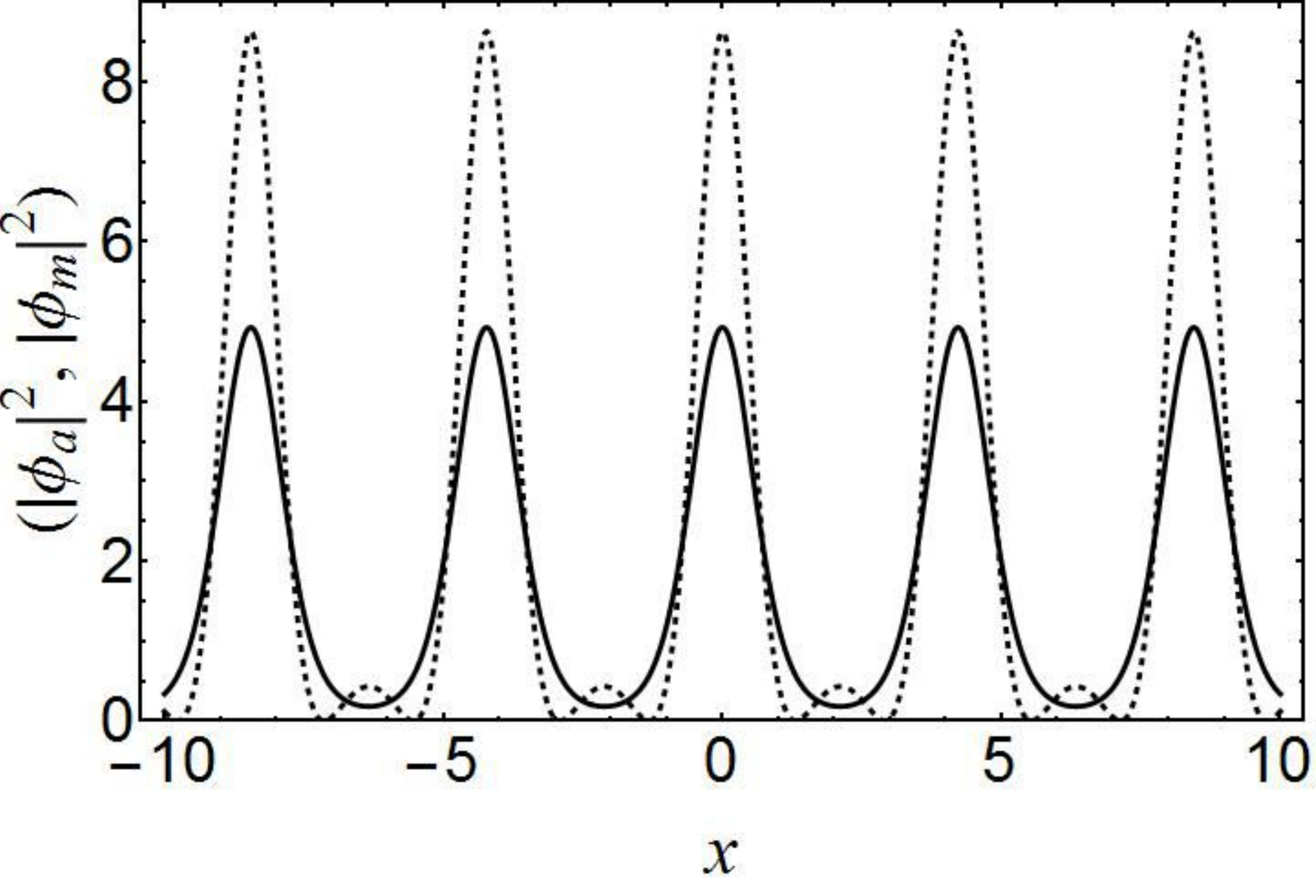}
\includegraphics[scale=0.28]{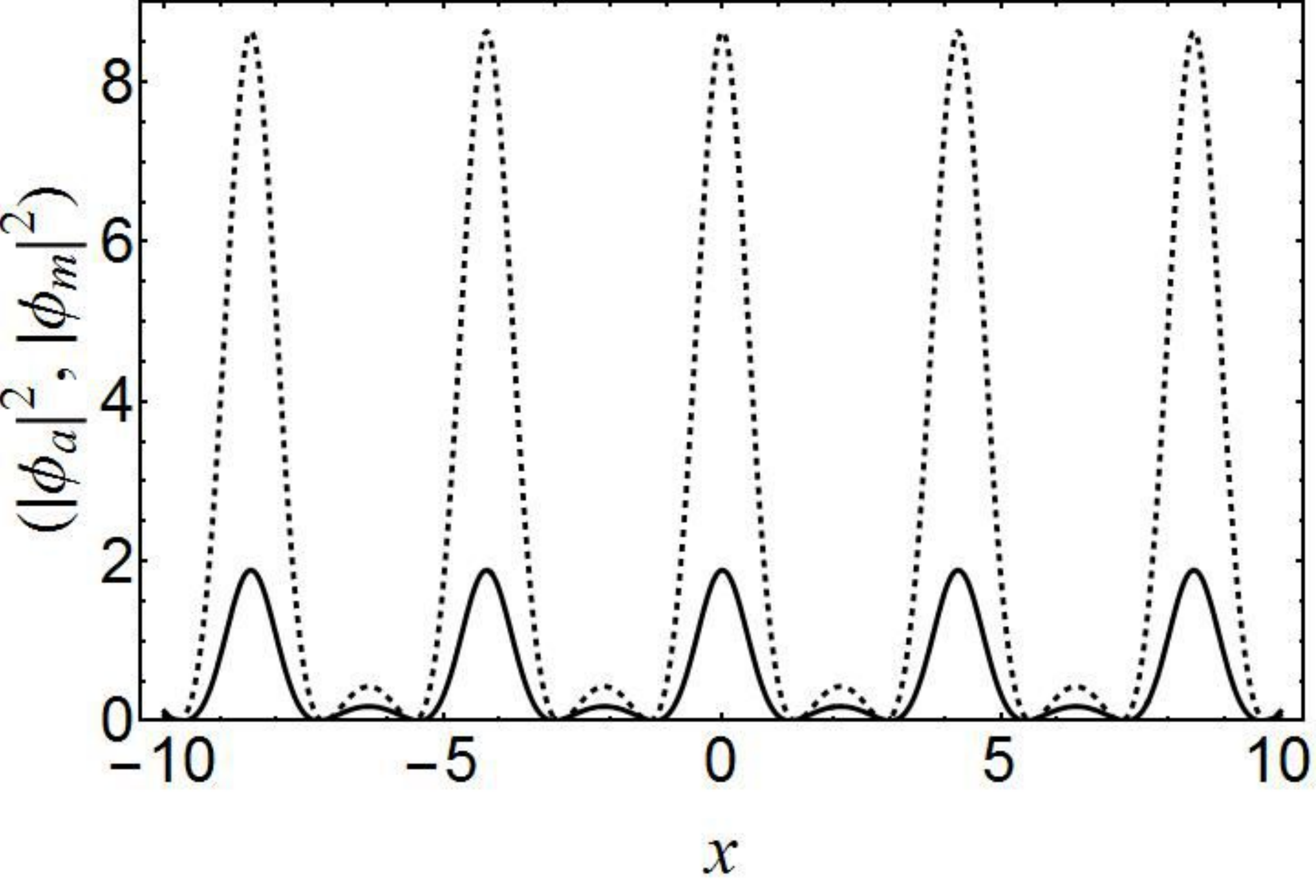}
\caption{Elevation and depression in density for static configuration. (left) shows elevation-elevation pair for atomic and molecular components. (right) shows elevation-depression pair for both the components.}
\label{fig4}
\end{center}
\end{figure}

\begin{figure}[t]
\begin{center}
\includegraphics[scale=0.39]{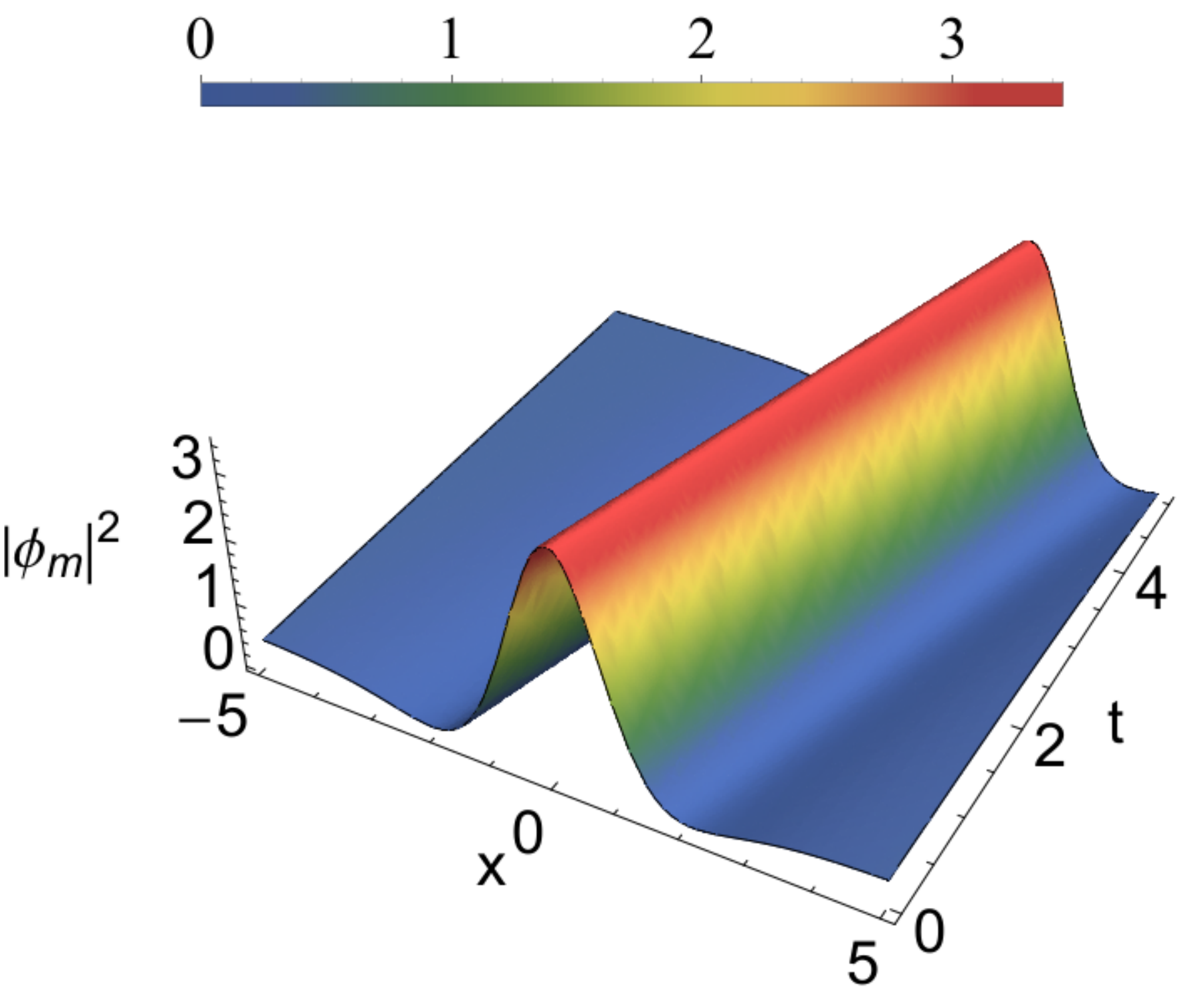}
\includegraphics[scale=0.39]{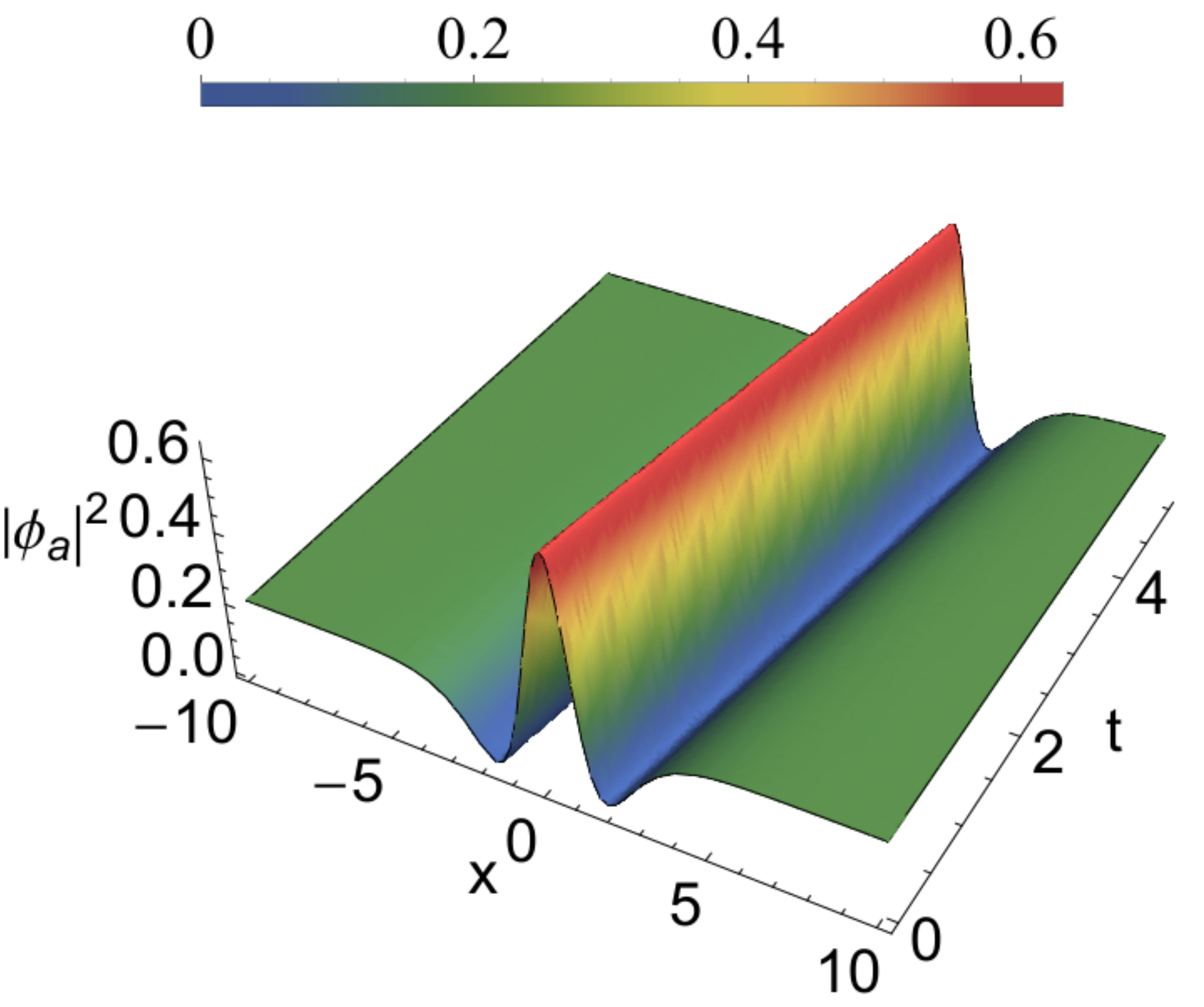}
\label{}
\end{center}
\end{figure}

\begin{figure}[t]
\begin{center}
\includegraphics[scale=0.3]{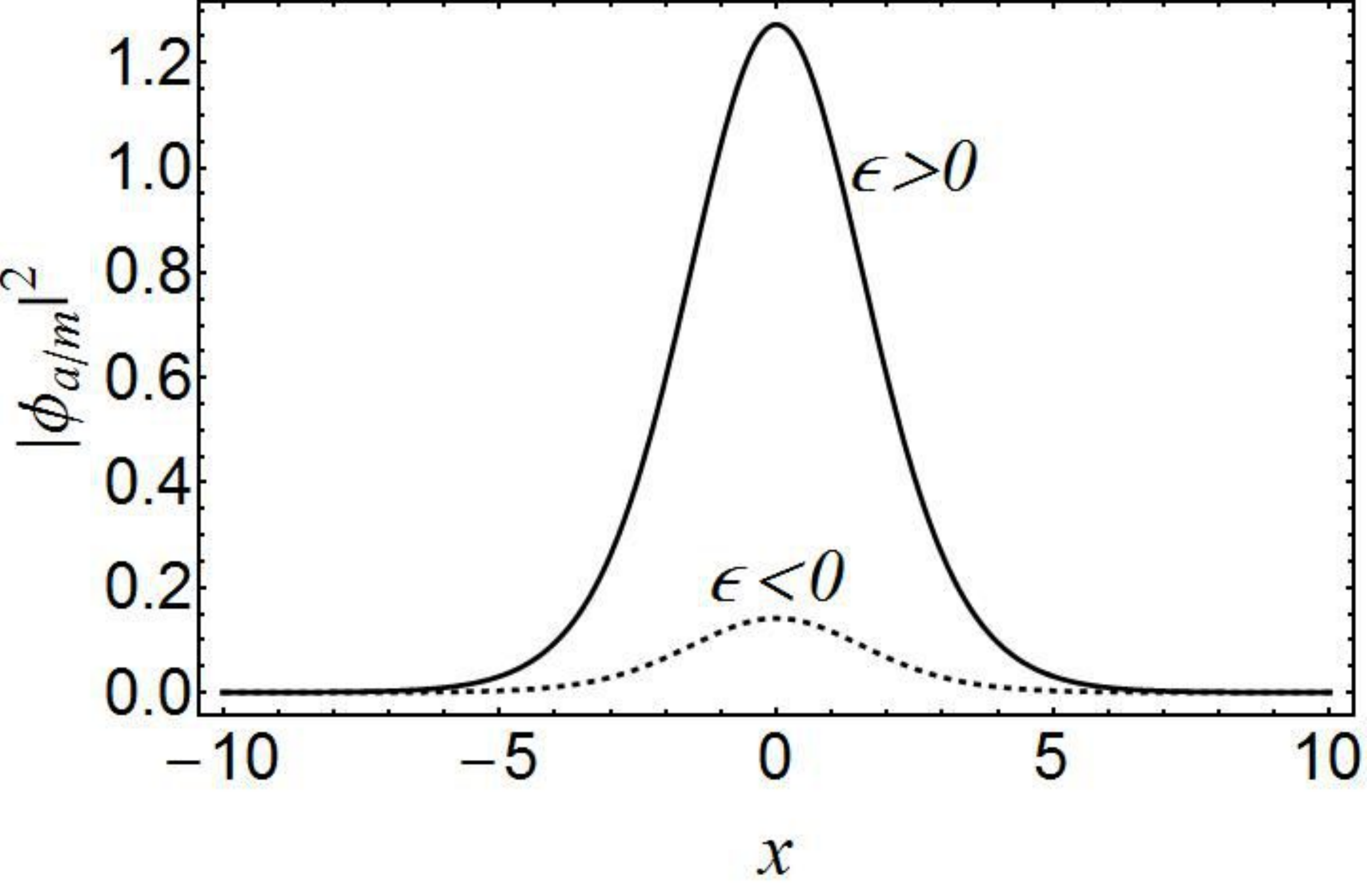}
\includegraphics[scale=0.3]{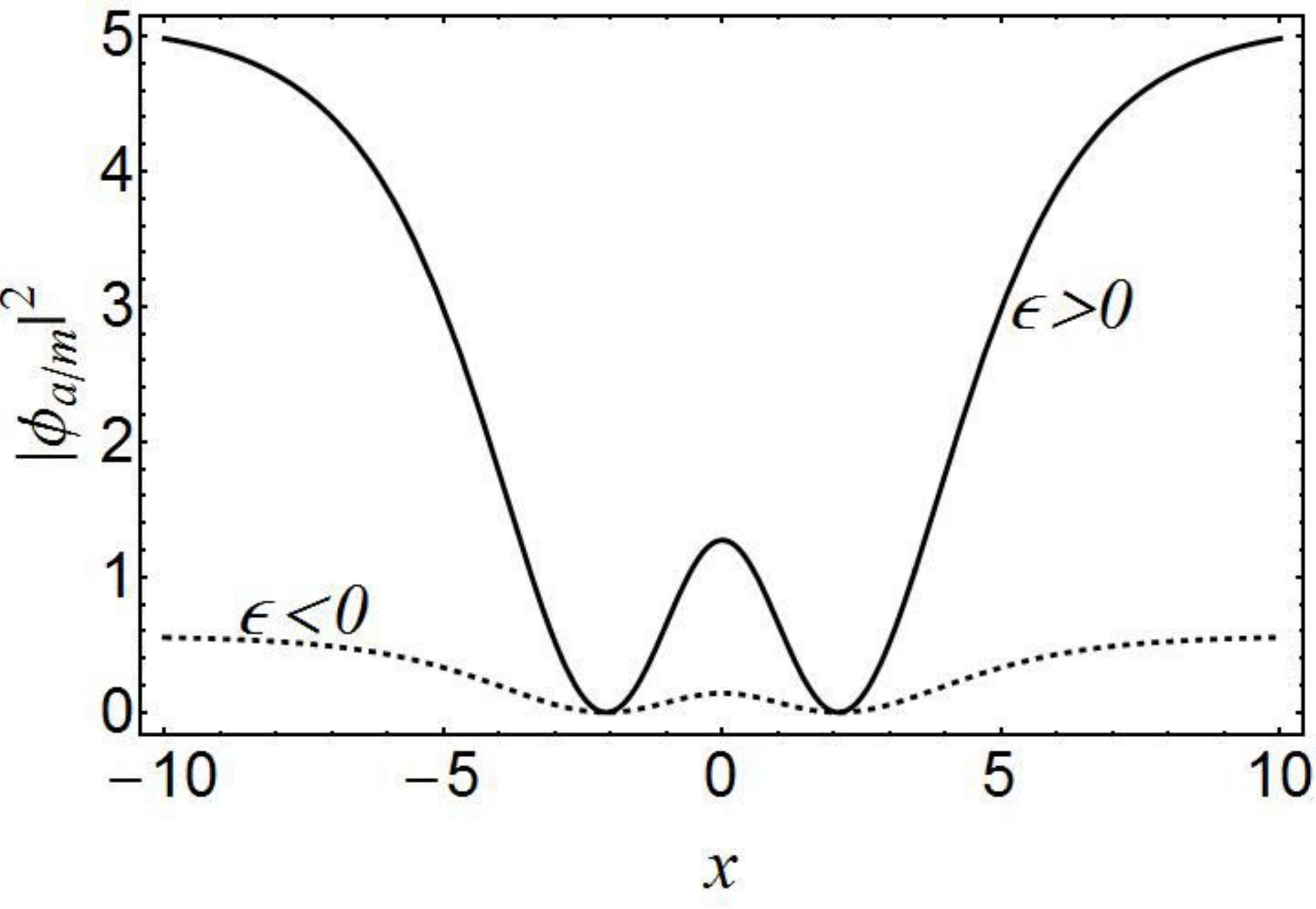}
\caption{Density profiles of localized solitons. (left bottom) shows density distribution for $\sigma=1$. (right bottom) shows W-type density distribution for $\sigma=3$ with a non-vanishing background.}
\label{fig5}
\end{center}
\end{figure}

\subsection*{Homo-density Gapped Solitons}
In addition to the \emph{cnoidal} waves, bright localized solitons for both the atomic and molecular components, are found as exact solutions:

\begin{eqnarray}
\phi_{a}(x,t) &=& \sigma_{0}\left(1 - \sigma\tanh^{2}\left[\beta(x-ut)\right]\right) e^{i (k x-\Omega t)} \label{sac} \\
\textrm{and\,\,\,\,\,}
\phi_{m}(x,t) &=& \sigma_{0}\left(1 - \sigma\tanh^{2}\left[\beta(x-ut)\right]\right) e^{2 i (k x-\Omega t)} \label{smc}
\end{eqnarray}
The mean field equations yield two distinct configurations
\begin{eqnarray}
\sigma_{0} &=&-\frac{\epsilon}{3\sqrt{2}\alpha};\beta^2=\frac{\epsilon}{3}, \\
\textrm{or\,\,\,\,\,}
\sigma_{0} &=& \frac{\epsilon}{\sqrt{2}\alpha}; ~\beta^2=-\frac{\epsilon}{3}
\end{eqnarray}

with $\Omega-k^{2}/2 = 2\epsilon/3$. The consistency conditions allow only two discrete values for the parameter $\sigma$: $\sigma = 1, 3$. Above solutions exist only for $g_{a} = g_{m} = -g_{am}$, i.e., when the self-interactions (atom-atom or molecule-molecule) are repulsive and the cross-interaction (atom-molecule) is attractive or vice versa. For $\epsilon>0$, $k^2=2\Omega-\frac{4}{3}\epsilon$, which is consistent with $\Omega>2\epsilon/3$, indicating a finite wave number is needed to excite solution for positive frequency case, while the negative half of the mismatch sets a lower limit for the frequency, $\Omega>-2\vert\epsilon\vert/3$. Density profiles of the localized solitons are depicted in Fig \ref{fig5}, showing distinct behaviour for the cases with and without background. Interestingly, for $\sigma=3$, we find a W-type soliton profile with background. On the other hand, for $\sigma=1$, one finds an asymptotically vanishing solitary excitation. 


\section*{Conclusion}
To summarize, we have investigated the reaction kinetics associated with distinct set of collective chemical waves with different density distributions. Photoassociation is found to dominantly regulate the rate of reactions and product formation. This can be used to produce selectively dense mixture of atoms and molecules. Possibility of formation of multidimensional spatio-temporal solitons in pure cubic media has been theoretically demonstrated previously, here we extend this prediction to matter-wave interactions in BEC systems where quadratic nonlinear  contribution due  to atom-molecule conversion is unavoidable. The results  obtained yield precise  conditions  under which two distinct pair of atomic-molecular BEC solitons can form, in terms of the parameters originating from atom-molecule coupling, atom-atom $s$-wave scattering, and the energy detuning between the atomic and molecular fields. The unique properties of ultracold energy regime lead to an effective quantization of scattering phase shift, enabling the interference between reaction pathways which contribute to the total reaction rate. Using this mechanism of controlling interference, one can switch the reaction on or off by
varying external fields. This new mechanism is a general property of ultracold reactions and will play a crucial role in their technological applications.

\end{document}